\documentclass{iopart}
\usepackage{iopams}  
\usepackage{a4wide}
\usepackage{amsfonts,latexsym}
\usepackage{graphicx,subfigure}
\newcommand{\be}{\begin{equation}}
\newcommand{\ee}{\end{equation}}
\newcommand{\bea}{\begin{eqnarray}}
\newcommand{\eea}{\end{eqnarray}}

\newcommand{\bel}[1]{\begin{equation}\label{#1}}
\newcommand{\beal}[1]{\begin{eqnarray}\label{#1}}

\begin{document}
%--------------------------------------------------------------------------------------

%--------------------------------------------------------------------------------------
\title[Bounce scenarios in the SVW generalization of the projectable Horava-Lifshitz gravity]{ Bounce scenarios in the Sotiriou-Visser-Weinfurtner generalization of the projectable Horava-Lifshitz gravity}
\author{E.~Czuchry}
\address{
Instytut
Problem\'ow J\c{a}drowych, ul. Ho\.za 69, 00-681 Warszawa, Poland}
\ead{eczuchry@fuw.edu.pl}

\begin{abstract}

The occurrence of a bounce in the FRW cosmology requires modifications of General
Relativity. An example of such a modification is the recently proposed
Ho\v{r}ava-Lifshitz theory of gravity, which includes a ``dark radiation'' term
with a negative coefficient in the analog of the Friedmann equation. A modification of the HL gravity, relaxing the "detailed balance" condition, brings additional terms to the equations of motion, corresponding to stiff matter. This paper presents comparison of the phase structure of the original and modified Ho\v{r}ava cosmology. Special attention is paid to the analysis of  a  wide range of bouncing solution, appearing in both versions of  the Ho\v{r}ava theory. 
\end{abstract}

%\maketitle

%--------------------------------------------------------------------------------------

%--------------------------------------------------------------------------------------
\section{Introduction}
%--------------------------------------------------------------------------------------

 There have been many attempts to modify
Einstein's theory of gravity to avoid an initial singularity. Some were made at the classical level,
some involve quantum effects. Examples include the ekpyrotic/cyclic model
(\cite{Ekp1,Ekp2,Cyclic1,Cyclic2,Cyclic3,pyrotech}) and loop quantum cosmology (\cite{ap1,ap2,ap3}), which
replace the Big Bang with a Big Bounce. Attempts to address these issues at
the classical level include braneworld scenarios (\cite{bw,rs}), where the
universe goes from an era of accelerated collapse to an expanding era without
any divergences nor singular behavior.  There are also higher order
gravitational theories and theories with scalar fields (see \cite{bc} for a
review of bouncing cosmologies). However it is fair to say that the issue of the
initial singularity still remains one of the key questions of the early Universe
cosmology.

Recently much effort has been devoted to studies of a proposal for a UV
complete theory of gravity due to Ho\v{r}ava~\cite{hor1,horava,Horava:2009if} and modifications of the theory \cite{horava,Nastase:2009nk, sfetsos,Sotiriou:2009,SVW} (for a recent review see \cite{mk}).  Because in the
UV the theory possesses a fixed point with an anisotropic, Lifshitz scaling
between time and space, this theory is referred to as the Ho\v{r}ava-Lifshitz
gravity. From the time at which the Ho\v{r}ava theory was presented, there is also quite much discussion of possible problems and
instabilities of Ho\v{r}ava-Lifshitz gravity
\cite{Charmousis:2009tc, LiPang,Bogdanos:2009uj,ak}. 
Numerous sophisticated versions contain new terms added to the original Lagrangian with attempt to make the proposal more general \cite{SVW}
and to solve the so called  strong coupling problem \cite{Charmousis:2009tc,sc1,sc2,blas,blas3}. Even so  it is still tempting to investigate issues opened by this theory and its modifications.

Soon after this theory was proposed many specific solutions of this
theory have been found, including cosmological ones
(\cite{hc1, hc2, Saridakis:2009bv, Brandenberger:2009yt,Nastase:2009nk, Mukohyama:2009zs, Minamitsuji:2009ii, Wang:2009rw,Takahashi:2009wc,lmp}).  It was also realized that the analog of the
Friedmann equation in the HL gravity contains a term which scales in the same way as
dark radiation in braneworld scenarios \cite {hc1,hc2,Saridakis:2009bv} and gives a negative
contribution to the energy density. Thus, at least in principle it is possible
to obtain non-singular cosmological evolution within Ho\v{r}ava theory, as it
was pointed out in \cite{hc1,Saridakis:2009bv,Brandenberger:2009yt,Saridakis:2009}. Such 
possibility  may have dramatic consequences for potential histories of the Universe --  other than avoiding the initial singularity.  New imaginable scenarios of cosmological evolution include contraction from the infinite size, bounce and then expansion to infinite size again, or eternal cycles of contraction, bounce and expansion.

Additional possibilities are brought by some interesting modifications modifications of HL gravity, either by softly breaking a detailed balance condition \cite{horava,Nastase:2009nk, sfetsos} or relaxing it completely \cite{Sotiriou:2009,SVW}. In the latter works Sotiriou, Visser and Weinfurtner  (SVW) in search for the more general renormalizable  gravitational theory took a gravitational action containing terms non only up to quadratic in curvature, like in original HL formulation, but also cubic ones, as suggested earlier in \cite{hc1,hc2}. Generalized Friedmann equation of this model include
 modified dark radiation term proportional to $\sim 1/a^4$   ($a$ is a scale factor) from the original HL formulation, and also additional $1/a^6$ term. This new term, negligible at large scales, becomes significant at small ones and modifies bounce solutions. Specially, with the opposite sign (the value of a coupling constant is arbitrary) than the $1/a^4$ term, it may compensate the dark radiation term at small distances and cancel the possibility of avoiding singularity, like in the HL gravity with the softly broken  detailed balance condition and negative spacial curvature \cite{SonKim}.
Thus one of the questions to be answered is how the additional terms in the generalized
Friedmann equations of the SVW HL gravity influence  the existence and stability of a
cosmological bounce.  

In this work we are going to investigate, with the help of the phase portrait techniques, how relaxing the detailed balance condition
affects the dynamics of the system, and then compare the results to those in the standard HL theory. We will we focus on non flat cosmologies, with space curvature $k=\pm1$,  allowing non-singular  solutions. Unlike in our previous paper \cite{ewa}, we are going to describe matter by
a cosmological stress-energy tensor  added to the gravitational field equations. Such analysis is more effective and avoids unneeded approximations and simplifications.
 In this hydrodynamical approach  two quantities:  density $\rho$ and  pressure $p$ describe matter properties. 
 
Nonetheless, constant parameter  $w$  of the equation of state is of course an idealization, hard to avoid at this level of research. It would be better to have history of the SVW HL Universe constructed in a similar way as in the standard   $\Lambda$CDM model, with  phases and epochs of different matter/radiation contents. Yet unless a rich structure of the original and the generalized HL theory, with additional coupling constants  whose range of values is not fully understood thus far, is investigated deeper, we shall use simpler tools.
 Thus in this paper, within  limited physical understanding of the theory and its parameters, we would rather present   lists of possibilities than likely physical solutions. With the progress of research in this field and better understanding of the nature of these parameters, it will be possible to assign more  physical interpretation to a set of solutions/scenarios found.

Related analysis of the generalized Ho\v{r}ava-Lifshitz cosmology have recently appeared in
\cite{ndb1} and \cite{ndb2}, which we become aware of while this work was being
typed. Those papers address a somewhat different set of issues, i.e. static solutions of the HL universe.
Here we are interested in stable and unstable solutions leading to cosmological bounce, and  focus on both cases of non flat universe ($k=-1$ and $k=1$) with the value of  a HL constant $\lambda$ arbitrary. We agree on the regions of overlap. General discussion of the full phase space of the original HL cosmology is contained e.g. in \cite{ps1,ps2}.

The structure of a paper is following:  in Section 2. we briefly sketch
the Ho\v{r}ava-Lifshitz gravity and cosmology. In Section 3. the possibility of
bounce in theory with detailed balance condition is discussed. In Section 4 we discuss phase portraits of the HL universe with condition of detailed balance relaxed.

%--------------------------------------------------------------------------------------
\section{Ho\v{r}ava-Lifshitz cosmology}
%--------------------------------------------------------------------------------------

The metric of Ho\v{r}ava-Lifshitz theory, due to  anisotropy in UV,  is written
in the $(3+1)$-dimensional ADM formalism:
\be
\label{eq1} ds^2=-N^2 dt^2 +g_{ij}(dx^i-N^idt)(dx^j-N^jdt),
\ee
where $N$, $N_i$ and $g_{ij}$ are dynamical variables.

%--------------------------------------------------------------------------------------
\subsection{Detailed balance}
%--------------------------------------------------------------------------------------

The action of Ho\v{r}ava-Lifshitz theory is~\cite{horava}
\bea
\label{eq3}
I &=& \int dt\, d^3x ({\cal L}_0 +{\cal L}_1), \label{action}\\
 {\cal L}_0 &=& \sqrt{g}N \left \{\frac{2}{\kappa^2}
(K_{ij}K^{ij}-\lambda K^2) +\frac{\kappa^2\mu^2 (\Lambda_W
R-3\Lambda_W^2)}{8(1-3\lambda)}\right \},  \nonumber \\
 {\cal L}_1  &=& \sqrt{g}N \left \{\frac{\kappa^2\mu^2(1-4\lambda)}{32(1-3\lambda)}R^2
-\frac{\kappa^2}{2\omega^4}Z_{ij} Z^{ij} \right\} ,\nonumber
\eea
where $K_{ij}=\frac1N \left[\frac12\dot g_{ij}-\nabla_{(i}N_{j)}\right]$ is extrinsic
curvature of  a space-like hypersurface with a fixed time, a dot denotes a
derivative with respect to the time $t$ and covariant
derivatives  are defined with respect to the spatial metric $g_{ij}$. Moreover
\be
Z_{ij}=C_{ij}-\frac{\mu\omega^2}{2}R_{ij} .
\ee
$\kappa^2$, $\lambda$, $\mu$, $\omega$ and $\Lambda_W$ are
constant parameters and the Cotton tensor, $C_{ij}$, is defined by
\be
\label{eq4} C^{ij}=\epsilon^{ikl} \nabla_k \left (R^j_{\
l}-\frac{1}{4}R\delta^j_l\right) = \epsilon^{ikl}\nabla_k R^j_{\ l}
-\frac{1}{4}\epsilon^{ikj}\partial_kR.
\ee
In (\ref{eq3}),  ${\cal L}_0$ is the kinetic part of the action, while ${\cal
  L}_1 $ gives the potential of the theory in the so-called
``detailed-balance" form.

Matter may be added by inserting a cosmological stress-energy tensor in gravitational field equation.  Within such framework we approximate the stress-energy tensor by two quantities: density $\rho$ and  pressure $p$, then simply add them to the vacuum equations by demanding the
correct limit as one approaches  General Relativity -- the low energy limit of the HL theory. Relation between $\rho$ and $p$ is given by the equation $p=w\rho$, with $w$ being the equation of state parameter.

Comparing the action  of the Ho\v{r}ava-Lifshitz theory in the IR limit to the Einstein-Hilbert
action of General Relativity, one can see that the speed of light $c$, Newton's
constant $G$ and the effective cosmological constant $\Lambda$
are
\be\label{eq5}
c=\frac{\kappa^2\mu}{4}\sqrt{\frac{\Lambda_W}{1-3\lambda}}, \ \
G=\frac{\kappa^2 c}{32\pi}, \ \  \Lambda=-\frac{3\kappa^4\mu^2}{3\lambda-1}\frac{\Lambda^2_W}{32}=\frac{3c^2}2 \Lambda_W,
\ee
respectively.  To have real value of speed of light $c$ emerging, the HL cosmological constant $\Lambda_W$ has to be negative for $\lambda>1/3$ and positive for $\lambda<1/3$. It is possible to obtain a positive cosmological constant $\Lambda_W$ in the IR limit $\lambda=1$ if one performs in (\ref{action}) an analytic continuation  of constant parameters $\mu \mapsto i\mu$ and $\omega^2\mapsto-i\omega^2$. 

The equations for Ho\v{r}ava-Lifshitz cosmology are obtained by  imposing conditions of
homogeneity and isotropy of the metric.  The associated ansatz is $ N=N(t)$, $N_i= 0$, 
$g_{ij}=a^2(t)\gamma_{ij} $ where  $a(t)$ is a scale factor and $\gamma_{ij}$ is a
maximally symmetric constant curvature metric, with a curvature $k=\{-1,0,1\}$. On this background 
\be
K_{ij}=\frac{H}{N}g_{ij}\,,\qquad R_{ij}=\frac{2k}{a^2}g_{ij}\,,\qquad C_{ij}=0\,,
\ee
where $H\equiv \dot a/a$ is the Hubble parameter.

The  gravitational action (\ref{action}) becomes:
\bea
S_{\rm FRW}=\int dt\, d^3x \, N a^3\,&&\left\{\frac{3(1-3\lambda)}{2\kappa^2}\frac{H^2}{N^2}+\frac{3\kappa^2\mu^2\Lambda_W}{4(1-3\lambda)}\left(\frac{k}{a^2}-\frac{\Lambda_W}{3}\right)
\right.\nonumber\\
&&
\left.-\frac{\kappa^2\mu^2}{8(1-3\lambda)}\frac{k^2}{a^4}\right\}.\label{hla3}
\eea
The equations of motion are obtained by varying the action (\ref{hla3}) with
respect to $N$, $a$ and $\varphi$,  setting $N = 1$ at the end of the
calculations  and adding terms with  density $\rho$ and pressure $p$,  leading to
\bea
H^2 &=&\frac{\kappa^2 \rho}{6(3\lambda-1)} \pm \frac{\kappa^4\mu^2}{8(3\lambda-1)^2} \left( \frac{ k\Lambda_W}{a^2} - \frac{\Lambda_W^2}2-\frac{ { k}^2 }{2a^4}\right)
, \label{hc1} \\
{\dot H}  &=& -\frac{\kappa^2(\rho+p)}{4(3\lambda-1)} \mp \frac{\kappa^4\mu^2}{8(3\lambda-1)^2}\left( \frac{ k\Lambda_W}{a^2} +  \frac{ { k}^2 }{4a^4} \right), \label{hc2}
\eea
and the  continuity equation:
\be
\dot{\rho}+3H(\rho+p)=0,\label{ce}
\ee
 The upper sign denotes the $\Lambda_W<0$ case, the lower one the analytic continuation $\mu\mapsto i\mu$ with a positive $\Lambda_W$.

The significant new terms in the above equations of motion are the
$(1/a^4)$-terms on the right-hand sides of (\ref{hc1}) and
(\ref{hc2}). They are reminiscent of the  dark radiation term in the braneworld cosmology \cite{BDEL}
and are present only if the spatial curvature of the metric is non-vanishing.

Values of constant parameters $\kappa^2$ and $\mu$ may be expressed in terms of cosmological constants according to (\ref{eq5}). We will also  work in units such that $8\pi G=1$ and $c=1$. Then
\be
\kappa^2=32\pi G c=4, \ \Lambda=\frac32 \Lambda_W,
\ee
and
\be
\frac{\mu^2}{1-3\lambda}=\pm\frac3{2\Lambda}.
\ee
Substituting the above expressions and the equation of state $p=w\rho$ to  (\ref{hc1}-\ref{hc2}) leads to
\bea
H^2 &=& \frac2{3\lambda-1}\left[\frac{\rho}3\pm\left(\frac\Lambda3-\frac{k}{a^2}+\frac{3}{4\Lambda}\frac{k^2}{a^4} \right)\right]\label{hc11}\\
\dot{H}&=&\frac{2}{3\lambda-1}\left[-\frac{(1+w)}2\rho\pm\left(\frac{k}{a^2}-\frac3{2\Lambda}\frac{k^2}{a^4}\right)\right].\label{hc22}
\eea

%--------------------------------------------------------------------------------------
\subsection{Beyond detailed balance}
%--------------------------------------------------------------------------------------
The gravitational action written in the "detailed balance" form (\ref{action}) (\cite{horava}) contains terms up to quadratic in the curvature. However the most general renormalizable theory contains also cubic terms, as it was pointed out in
\cite{hc1,hc2}.  Thus Sotiriou, Visser and Weinfurtner  (\cite{Sotiriou:2009,SVW})  built a theory with projectability condition $N=N(t)$, as in original Ho\v{r}ava theory, but without the detailed balance condition.
This  led to Friedmann equations with an additional term $\sim1/a^6$ and uncoupled coefficients:
 \bea
H^2 &=& \frac2{(3\lambda-1)}\left(\frac{\rho}3 + \sigma_1 + \sigma_2  \frac{k}{a^2} + \sigma_3 \frac{k^2}{a^4} + \sigma_4 \frac{k}{a^6}\right),\\
\dot{H}&=& \frac2{(3\lambda-1)}\left(-\frac p2 - \frac\rho2 - \sigma_2\frac{k}{ a^2} -2\sigma_3\frac{k^2}{ a^4} -3\sigma_4\frac{k}{ a^6} \right).\
\eea
Values of constants $\sigma_3$, $\sigma_4$ are arbitrary. In order to coincide with the Friedmann equations in the IR limit $\lambda=1$ and for large $a$ (terms proportional to  $1/a^4$ and  to $1/a^6$ are then neglible) one has to set $\sigma_1=\Lambda/3$ and $\sigma_2=-1$. Thus the above equations take the following forms:
 \bea
H^2 &=& \frac2{(3\lambda-1)}\left(\frac{\rho}3 + \frac\Lambda{3} -  \frac{k}{a^2} + \sigma_3 \frac{k^2}{a^4} + \sigma_4 \frac{k}{a^6}\right)\label{h2ndb},\\
\dot{H}&=& \frac2{(3\lambda-1)}\left(-\frac{\rho(1+w)}2 + \frac{k}{ a^2} -2\sigma_3\frac{k^2}{ a^4} -3\sigma_4\frac{k}{ a^6} \right)\label{hdndb},\
\eea
where we have used the equation of state $p=w\rho$. Density parameter follows the standard evolution equation (\ref{ce}).
New terms, proportional to $1/a^6$, appearing in the analogs of Friedmann equations, mimic stiff matter (e.g. such that $\rho=p$ and $\rho_{\textrm{stiff}}\sim1/a^6$). These terms are negligibly small at large scales, but may play a significant role at small values of a scale parameter.

%--------------------------------------------------------------------------------------
\section{Bounce stability in the original HL theory}
%--------------------------------------------------------------------------------------
In order to investigate the appearance of a bounce in the original HL gravity, we are going  first to simplify the equations of motion  (\ref{hc1}-\ref{hc2}) and then to reduce them with respect to  (\ref{hc1}). In this way we will obtain the two dimensional dynamical system describing the evolution of  $a$ and $H$.

Solving  Eq. (\ref{hc11}) for $\rho$ gives
\be
\rho=\frac{3(3\lambda -1)}2H^2\mp\left(\Lambda-3\frac{k}{a^2}+\frac9{4\Lambda}\frac{k^2}{a^4}\right).
\ee
Inserting the above formula to  (\ref{hc22}) leads to
\be
\dot{H}=\frac{\pm1}{3\lambda-1}\left[\left(1+w\right)\Lambda-\left(3w+1\right)\frac{k}{a^2}+\frac{3\left(3w-1\right)}{4\Lambda}\frac{k^2}{a^4}\right]-\frac32\left(1+w\right)H^2\label{pe1}.
\ee
Equation (\ref{pe1}) and  the definition of the Hubble parameter:  
\be
\dot{a}=aH\label{pe2},
\ee  
provide the two dimensional dynamical system for variables $a$ and $H$.

To find the finite critical points we set all right-hand-sides of
equations (\ref{pe1}-\ref{pe2}) to zero. This
gives two points:
\bea
P_1:& a^2=\frac{3k}{2\Lambda},& H=0,\\
P_2:& a^2=\frac{(3w-1)k}{(1+w)2\Lambda},&H=0.
\eea
These points are finite, unless $w=-1$. In the latter case point $P_2$ is moved to infinity. Point $P_1$  exists for $k/\Lambda>0$. Point $P_2$  exists for $w>1/3$ and $k/\Lambda>0$ or $w<1/3$ and $k/\Lambda<0$. Thus those two points exist both at the same time for $w>1/3$.

 Stability properties of the critical points are
determined by the eigenvalues of the Jacobian of the system
(\ref{pe1}-\ref{pe2}). More precisely, one has to linearize
transformed equations (\ref{pe1}-\ref{pe2}) at each point. Inserting
$\vec{x}=\vec{x}_0+\delta\vec{ x}$, where $\vec{x}=(a,H)$, and keeping terms
up to 1st order in $\delta\vec{x} $ leads to an evolution equation of the form
$\delta\dot{\vec{ x}}=A\delta\vec{x}$. Eigenvalues of $A$ describe stability
properties at the given point. Critical points at which all the eigenvalues have
real parts different from zero are called hyperbolic. Among them  one can distinguish  sources (unstable) for positive real parts, saddle for
real parts of different sign and sinks (stable) for negative real parts. If at least one eigenvalue has a
zero real part (non-hyperbolic critical point) it is not
possible to obtain conclusive information about the stability
from just linearization and needs to resort to other tools like
e.g. numerical simulation \cite{xx}. 

Eigenvalues  at $P_1$  are following:
$$\left(-2 \sqrt{\frac{\mp2\Lambda}{3(1-3\lambda)}}, 2\sqrt{\frac{\mp2\Lambda}{3(1-3\lambda)}}\right),$$
For all admitted values of $\Lambda$ and $\lambda$, expression $\mp\Lambda/(1-3\lambda)$ is negative, thus $P_1$ is a center (both eigenvalues of $A$ are purely imaginary at this point).

 Eigenvalues at the second  finite critical  point $P_2$ read as:
$$\left( -2 \sqrt\frac{\mp2\Lambda (1 + w)}{(1-3\lambda)(1- 3 w)},2 \sqrt\frac{\mp2\Lambda (1 + w)}{(1-3\lambda)(1- 3 w)}\right).$$
Depending on the value of parameter $w$ the point $P_2$ may be  a center  or  a saddle (two real numbers with opposite signs). 
Precisely, $P_2$ is a  linear center (non-hyperbolic center with one eigenvector) for $w=-1$ ($k/\Lambda<0$),  a center for $-1<w<1/3$ ($k/\Lambda<0$)  and  a saddle for $w>1/3$ ($k/\Lambda>0$).

Properties of the critical points $P_1$ and $P_2$ in dependence on the values of  $\Lambda$, $k$, $w$ are gathered in the Table 1.
$\Lambda<0$ corresponds to solutions of (\ref{pe1})-(\ref{pe2}) with the upper sign, the case of $\Lambda>0$ to  the lower sign in (\ref{pe1})-(\ref{pe2}).
\\

\begin{table}[!h]
\begin{center}
\begin{tabular}{|c|c|c|c|c|c|}
\hline
    $k/\Lambda$ & $w$ &$P_{1}$ & stability&$P_{2}$ & stability  \\
\hline
\hline
    & $>\frac13$  & +& center&+&saddle\\
   \cline{2-6}
   $>0$&$-1<w<\frac13$&+&center&-&\\
    \cline{2-6}
   &$-1$&+&center&-&\\   
      \cline{1-6}
& $>\frac13$&-&&-&\\
    \cline{2-6}
     $<0$& $-1<w<\frac13$&-&&+&center\\
      \cline{2-6}
     & $-1$&-&&moves to $\infty$&linear center\\
\hline
   \end{tabular}\caption{Properties of finite critical points in the HL theory. The plus sign stands for ``exists'' and the minus sign stands for ``does not exists''.} 
\end{center}
\end{table}

 To find critical points that occur at infinite values of the parameters we
rescale the infinite space  $(a,H)$ into a finite Poincar\'{e} sphere (as in \cite{frolov, frolov2}) in such a way that the new coordinates $(\tilde{a},\tilde{H})$ are written in polar coordinates ${r,\phi}$: $\tilde{a}=r\cos\phi$ and $\tilde{H}=r\sin\phi$ and:
\bea
a &=& {{r} \over 1-r} \cos\phi , \label{poin1} \\
H &=& {{r } \over 1-r}\sin\phi,  \label{poin2} 
\eea
We also  rescale the
time parameter  $t$ by defining the new time parameter $T$ such that: $d{T} = dt/(1-r)$. In these
coordinates our phase space  is contained within a sphere of radius one
-- infinity corresponds to $r=1$. More precisely, semi-sphere, as a scale factor $a$ may take only nonnegative values. 

This leads to the dynamical equations in terms of $r$, $\phi$ and their derivatives with respect to new time $T$. 
Taking limit $r = 1$ we obtain:
\bea
r'(T)&=&0,\\
\phi'(T)&=&- \frac{5 + 3 w}{2}  \cos\phi \sin^2\phi.
\eea
Putting r.h.s. of the above equations to zero, we find 4 solutions:
\bea
P_3&=& (1,0)\nonumber\\
P_4&=& (1,\pi/2)\nonumber\\
P_5&=& (1,\pi)\nonumber\\
P_6&=& (1,3\pi/2)\nonumber
\eea
in polar coordinates $(r,\phi)$.  Point $P_5$ is nonphysical (a negative $a$) and shall be removed from further discussions. Eigenvalues of the Jacobian matrix at the above points are following:
\bea
&& (0,0)\ \textrm{at}\ P_3:\nonumber\\
&& \left(\frac{5 + 3 w}2, 3\frac{1 +  w}2\right)\ \textrm{at}\ P_4\nonumber\\
&& \left( -\frac{5 + 3 w}2,  -3\frac{1 +  w}2\right)\ \textrm{at}\ P_6\nonumber
\eea
The point $P_3$ is non-hyperbolic and we determine its properties by numerical simulations for each set of parameters. Unless $w=-1$ points $P_4$ and $P_6$ are respectively a repelling  and an attracting node. For $w=-1$ the  finite 
fixed point $P_2$ is moved to $(\infty,0)$  becoming $P_3$, which is then a linear center. For this value of $w$ points $P_4$ and $P_6$ are non-hyperbolic.  It follows from numerical simulations that they are  saddles then and ends of a separatrice. 

Fig. 1 shows the phase portrait of HL universe containing matter with equation of state parameter $w>1/3$, $k/\Lambda>0$, Figure 2. shows phase portrait for $-1<w<1/3$, $k/\Lambda<0$ and Fig. 3 for $w=-1$, $k/\Lambda<0$. One has to note that these figures contain the deformed phase space, scaled to fit on the finite Poincar\'{e} sphere. One may have the impression that they describe regions in which e.g. the scale factor $a$ increases although the Hubble parameter $H$ is negative. However  it is the parameter $\tilde{a}$ that is increasing on the diagram, not the scale factor $a$.

\begin{figure}
\begin{center}
\includegraphics[height=100mm]{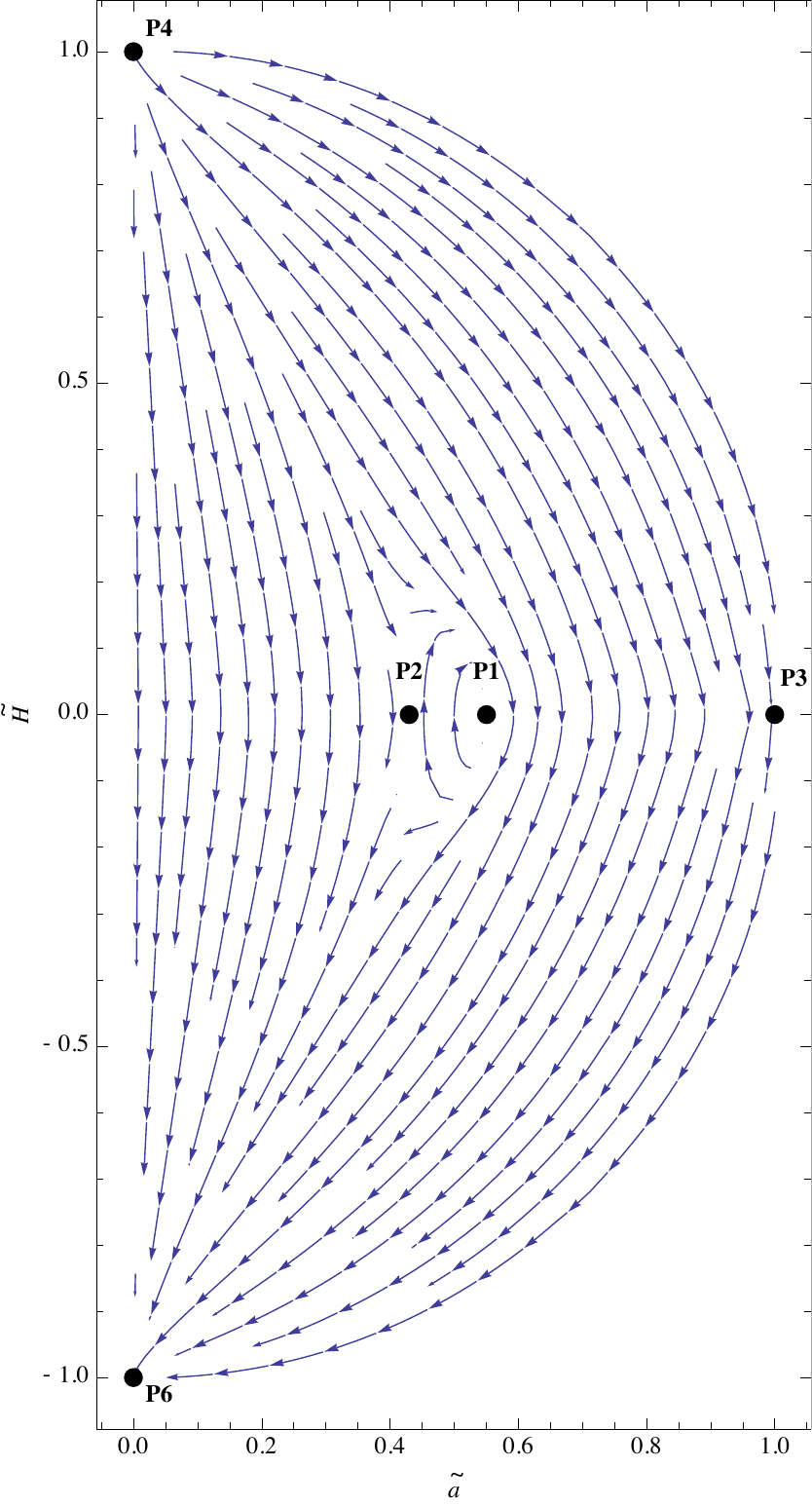}
\caption{Projected phase space of HL universe with $k\Lambda>0$ and $w>1/3$.}
\end{center}
\end{figure}
\begin{figure}
\begin{center}
\includegraphics[height=100mm]{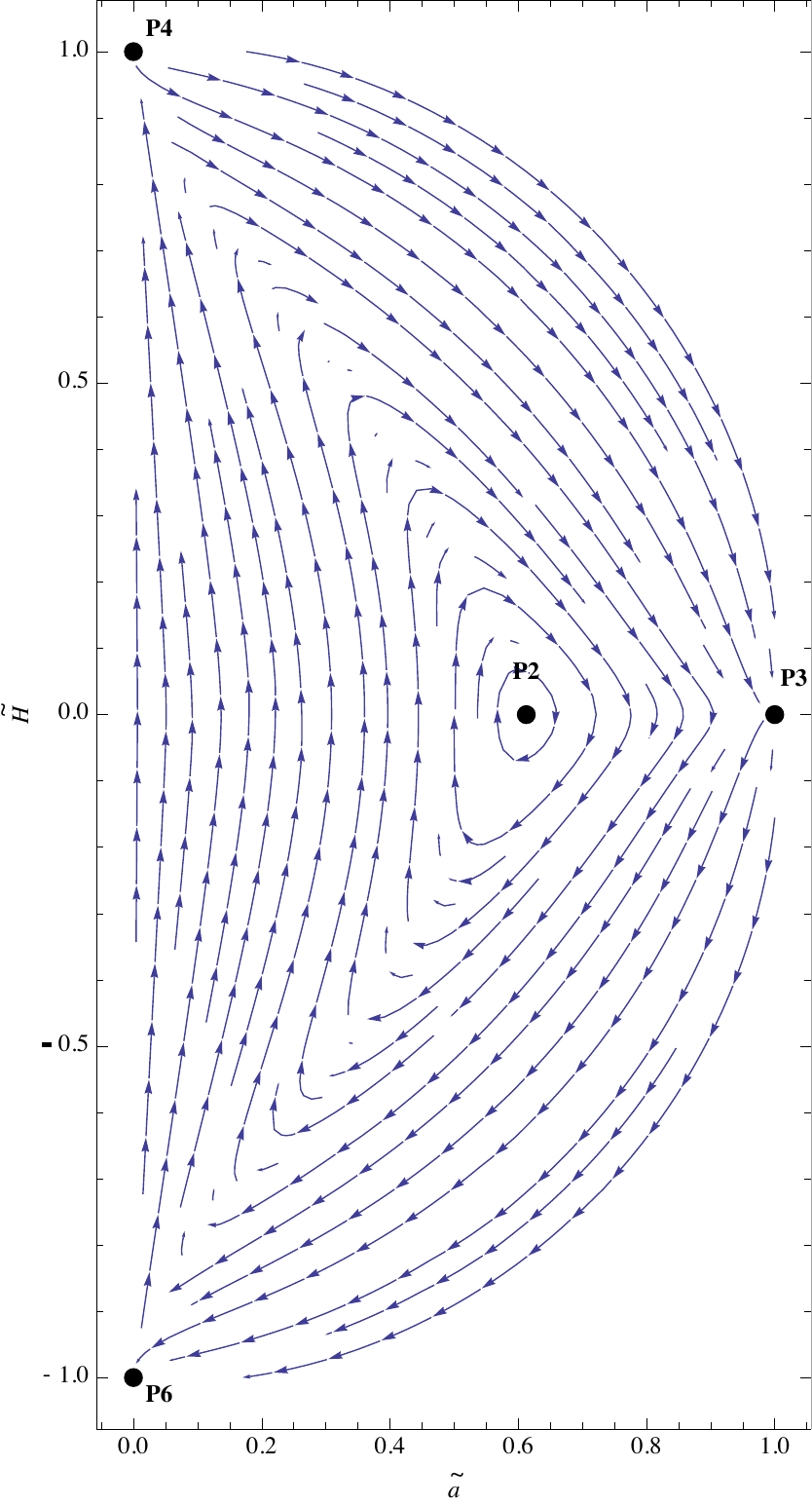}
\caption{Projected phase space of HL universe with $k/\Lambda<0$ and $-1<w<1/3$.}
\end{center}
\end{figure}

\begin{figure}
\begin{center}
\includegraphics[height=100mm]{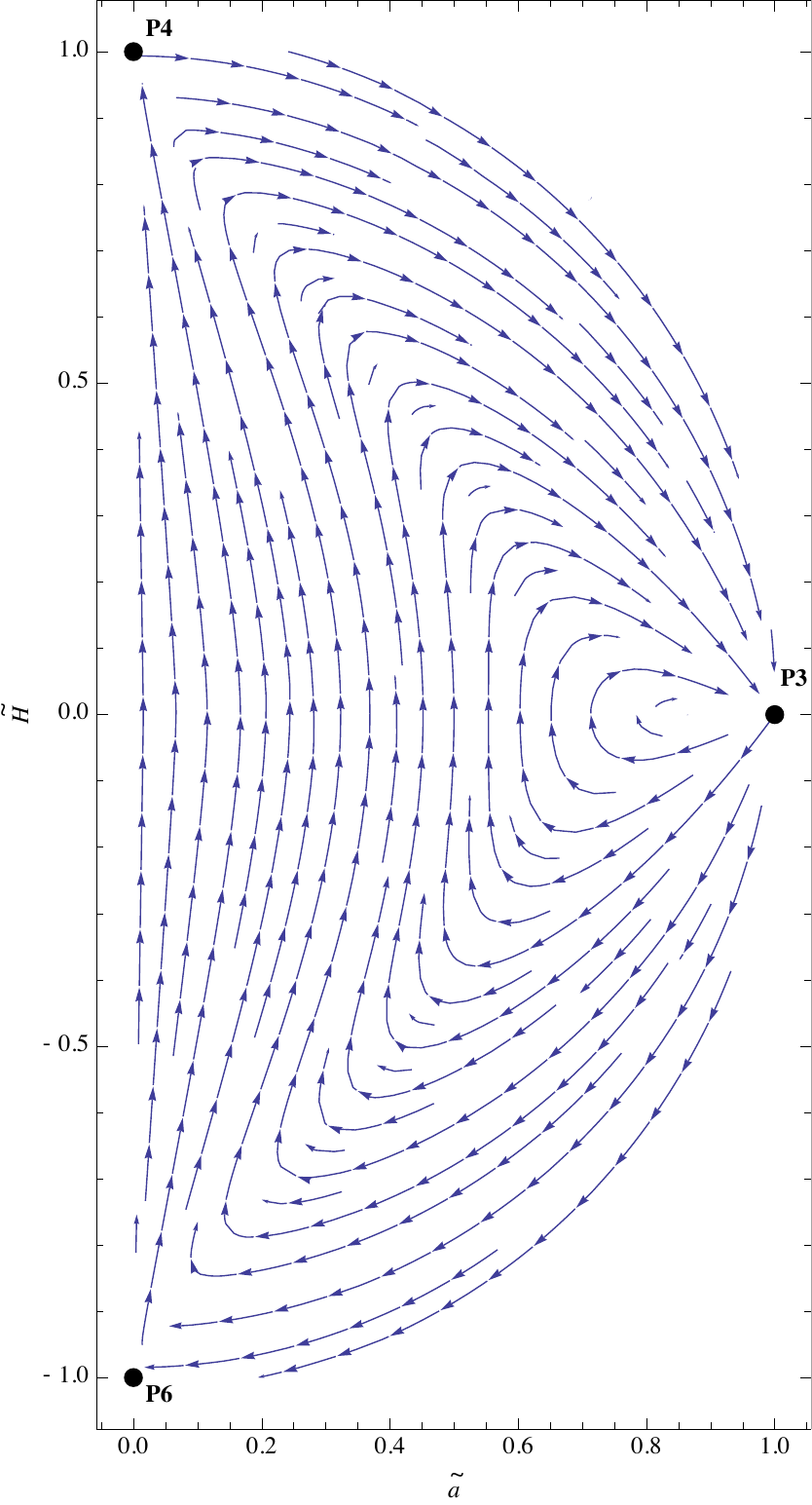}
\caption{Projected phase space of HL universe with $k/\Lambda<0$ and $w=-1$.}
\end{center}
\end{figure}

Bounce scenarios are thus possible when critical points exists. If these point  are centers then there are closed orbits around them and the Universe goes through eternal oscillations: expansion, collapse to a finite size, expansion etc. Point $P_{1}$ may be a center (for certain values of parameters), but  then $\rho=0$, which is  physically not interesting. More interesting case is when  $P_{2}$ is a center, then there are  closed orbits  with a non-zero density $\rho$. The third bounce scenario is around the  linear center $P_{2}$ (moved to $\infty$ and coinciding with $P_{3}$). In this case there are onefold closed trajectories: universe starts from a static one ($H=0$) and infinite ($a=\infty$), goes through a period of collapsing to a finite size, then after a bounce starts expansion and  finishes as a static infinite universe.

%--------------------------------------------------------------------------------------
\section{Bounce stability in the SVW generalization}
%--------------------------------------------------------------------------------------
Relaxing detailed balance condition leads to the generalized Friedmann equations (\ref{h2ndb}-\ref{hdndb}) with additional term $\sim1/a^6$ and uncoupled coefficients.

We may solve eq. (\ref{h2ndb}) for $\rho$ obtaining:
\be
\rho=3 \frac{(3\lambda-1)}2 H^2 -\Lambda - 3\frac{k}{ a^2} -3  \frac{\sigma_3k^2}{ a^4} -   \frac{3\sigma_4k}{a^6}.
\ee
Substituting this expression on $\rho$ into (\ref{hdndb}) and using the equation of state $p=w\rho$ leads to
\bea
\dot{H}&=&  \frac2{3\lambda-1}\left(   \frac{\Lambda(1 + w)}{2} -  \frac{ k(1+3w)}{2 a^2} \right.\nonumber \\&+& \left.\frac{\sigma_3 (-1 + 3 w)k^2}{2 a^4}
+\frac{ 3\sigma_4 (1 +  w)k}{2 a^6}\right)-\frac{3(1 + w)}{2} H^2 \label{doth}.
\eea
The above equation, together with the definition of the Hubble parameter provides the two dimensional dynamical system for variables $a$ and $H$.

Finite critical points are solutions of the  equations (\ref{pe2}) and (\ref{doth}) with r.h.s. set to zero. Hence these points fulfill $H=0$ and:
\be
 \Lambda(1 + w)a^6 -  { k(1+3w)}{ a^4} + {\sigma_3 (-1 + 3 w)k^2}{ a^2}
+{ 3\sigma_4 (-1 +  w)k}=0\label{bic}
\ee
The latter one is a bicubic equation, which may be simplified in few special cases.

\subsection{Cosmological constant $w=-1$}

 For $w=-1$ equation (\ref{bic}) reduces to a biquadratic equation:
 \be
  k a^4 -2\sigma_3 k^2{ a^2}
- 3\sigma_4 k=0\label{biq}.
\ee
Solutions of (\ref{biq}) are following:
\bea
P_{1}:&&a^2=k\sigma_3-\sqrt{\sigma_3^2+3\sigma_4},\\
P_{2}:&& a^2=k\sigma_3+\sqrt{\sigma_3^2+3\sigma_4}.
\eea
Point $P_{1}$ exists when $\{ (k\sigma_3>0, \sigma_4<0);  (|\sigma_4|<\sigma_3^2/3)\}$.
Point $P_{2}$ exists for:
\be\left\{(k\sigma_3>0, \sigma_4>0);\ (k\sigma_3>0, \sigma_4<0, |\sigma_4|<\sigma_3^2/3);\ (k\sigma_3<0, \sigma_4>0)\right\}.
\ee  
Stability properties of the critical points found are given by eigenvalues of the Jacobian matrix $A$ of the system (\ref{pe2}),(\ref{doth}).
Eigenvalues of $A$ at $P_{1}$ are following:
\be
\left(\pm\sqrt{\frac{-k C_1}{3\lambda-1}},\mp\sqrt{\frac{-k C_1}{3\lambda-1}}\right) ,
\ee
where $C_1$ denotes expression in $k,\sigma_3,\sigma_4$, positive when the point $P_{1}$ exists. Similarly, eigenvalues of $A$ at the point 
$P_{2}$ read as:
\be
\left(\pm\sqrt{\frac{k C_2}{3\lambda-1}},\mp\sqrt{\frac{k C_2}{3\lambda-1}}\right) ,
\ee
where $C_2$ denotes expression in $k,\sigma_3,\sigma_4$  being positive when the point $P_{2}$ exist. 

Thus for $k/(3\lambda-1)>0$ the point $P_{1}$ is a center and $P_{2}$  an unstable saddle,  for 
$k/(3\lambda-1)<0$  $P_{1}$ is a saddle and $P_{2}$  a center -- provided that the values of  $k,\sigma_3,\sigma_4$ allow their physical existence ($a^2>0$).  

Density $\rho$ at those points is equal to 
\be
\rho=-\frac{2 \sigma_3^3+9 \sigma_3 \sigma_3+9 L \sigma_4^2\pm 2 k \sigma_3^2 \sqrt{\sigma_3^2+3 \sigma_4}\pm 6 k \sigma_4 \sqrt{\sigma_3^2+3 \sigma_4}}{9 \sigma_4^2},
\ee
where '+' corresponds to the point $P_{1}$ and '-' to $P_{2}$. Thus at $P_{1}$ density $\rho>0$ if this point exists and $(k<0, \sigma_3<0, 0>\Lambda>1/\sigma_3)$. At the point $P_{2}$ density is positive if
 $(k<0,\sigma_3<0,\Lambda<0)$.

\subsection{Radiation $w=1/3$}

When $w=1/3$ equation (\ref{bic}) reduces to the following one:
\be
\frac{2\Lambda}3 x^3-kx^2-k\sigma_4=0\label{biq1},
\ee
where $x=a^2$.  The discriminant of the cubic polynomial $a_3y^3+a_2y^2+a_1y+a_0$
is of the following form:  $\Delta=18a_0a_1a_2a_3-4a_2^3a_0+a_2^2a_1^2-4a_3a_1^3-27a_3^2a_0^3$.
 Discriminant of the  equation (\ref{biq1}) reads as:
\be
\Delta=-4k^2\sigma_4(k^2+3\Lambda^2\sigma_4).
\ee

For $\Delta>0$ the cubic equation (\ref{biq1})  has three real solutions. Condition $\Delta>0$ is fulfilled for a nonflat universe ($k\neq0$) when $\sigma_4<0$ and $|\sigma_4|<1/(3\Lambda^2)$.
Otherwise (\ref{biq1}) -- the equation with real coefficients -- has one real solution and two nonreal complex conjugate roots ($\Delta<0$) or multiple real roots ($\Delta=0$). 

Physical points exist if  real solutions $x=a^2>0$. Equation (\ref{biq1}) cannot have three real positive roots, as it is implied by Vi\`ete's formulas -- precisely, by the second formula of the following ones:
\bea
\frac{3k}{2\Lambda}&=&x_1+x_2+x_3,\\
0&=&x_1x_2+x_2x_3+x_3x_1,\\
\frac{3k\sigma_4}{2\Lambda}&=&x_1x_2x_3.
\eea

If there are three real solutions ($\Delta>0$), one or two of them may be also positive. There is one positive solution if $k/\Lambda<0$ and two positive solution if  $k/\Lambda>0$.

Multiple real solutions exist when $\Delta=0$ hence when $\sigma_4=0$ or $\sigma_4=-k^2/(3\Lambda^2)$. The former case corresponds to HL theory with the detailed balance condition, the latter case implies  solutions: 
\bea
Q_{1}:\ && a^2=-k/(2\Lambda),\\
Q_{2}:\ && a^2=k/\Lambda\ \ \textrm{(double root)}, 
\eea
and $H=0$. Depending on the sign of $k/\Lambda$ one of the two solutions has physical meaning.

One real solution ($\Delta<0$), i.e. when $\sigma_4>0$ or $\sigma_4<-1/(3\Lambda^2$), has positive value if $k\sigma_4/\Lambda>0$.

Eigenvalues of the Jacobian matrix at the critical points $a_x$ are following:
\be
\left(-\frac{2}{a_x^3}\sqrt{\frac{k(3\sigma_4+a_x^4)}{3\lambda-1}},\frac{2}{a_x^3}\sqrt{\frac{k(3\sigma_4+a_x^4)}{3\lambda-1}}\right).
\ee
Thus they may be saddles or centers, depending on the sign of $k/(3\lambda-1)$ and $3\sigma_4+a_x^4$. 

We  can describe more precisely the case when $\sigma_4=-k^2/(3\Lambda^2)$ ($\sigma_4=0$ case is described within original HL cosmology) and critical points are $a_x^2=-k/(2\Lambda)$ ($Q_{1}$) or $a_x^2=k/\Lambda$ ($Q_{2}$). Then the eigenvalues of Jacobian matrix read as:
\be
\left(2\sqrt{6}\sqrt{\frac{\Lambda}{3\lambda-1}},-2\sqrt{6}\sqrt{\frac{\Lambda}{3\lambda-1}}\right)\ \ {\textrm{at}} \ \  Q_{1},
\ee
 and
\be
\left(0,0\right)\ \ {\textrm{at}} \ \  Q_{2}.
\ee
Therefore the point $Q_{1}$ may be  a saddle or a center, depending on the sign of $\Lambda/(3\lambda-1)$.  The points  $Q_{2}$ is non-hyperbolic, numerical simulations (Fig. 5) show that it is a  cusp. Comparing the neighborhood of $Q_2$ on both this figures, one can see the difference between the deformed phase space and the non-deformed one.
At the former one there are regions in which the parameter $\tilde{a}$ increases although the  parameter $\tilde{H}$ is negative, whereas this behavior is absent at the latter one.
This is due to the fact that parameters $\tilde{a}$ and $\tilde{H}$ are geometric objects without the same physical meaning as the scale factor $a$ and the Hubble parameter $H$.

\begin{figure}\label{xx}
\begin{center}
\includegraphics[height=100mm]{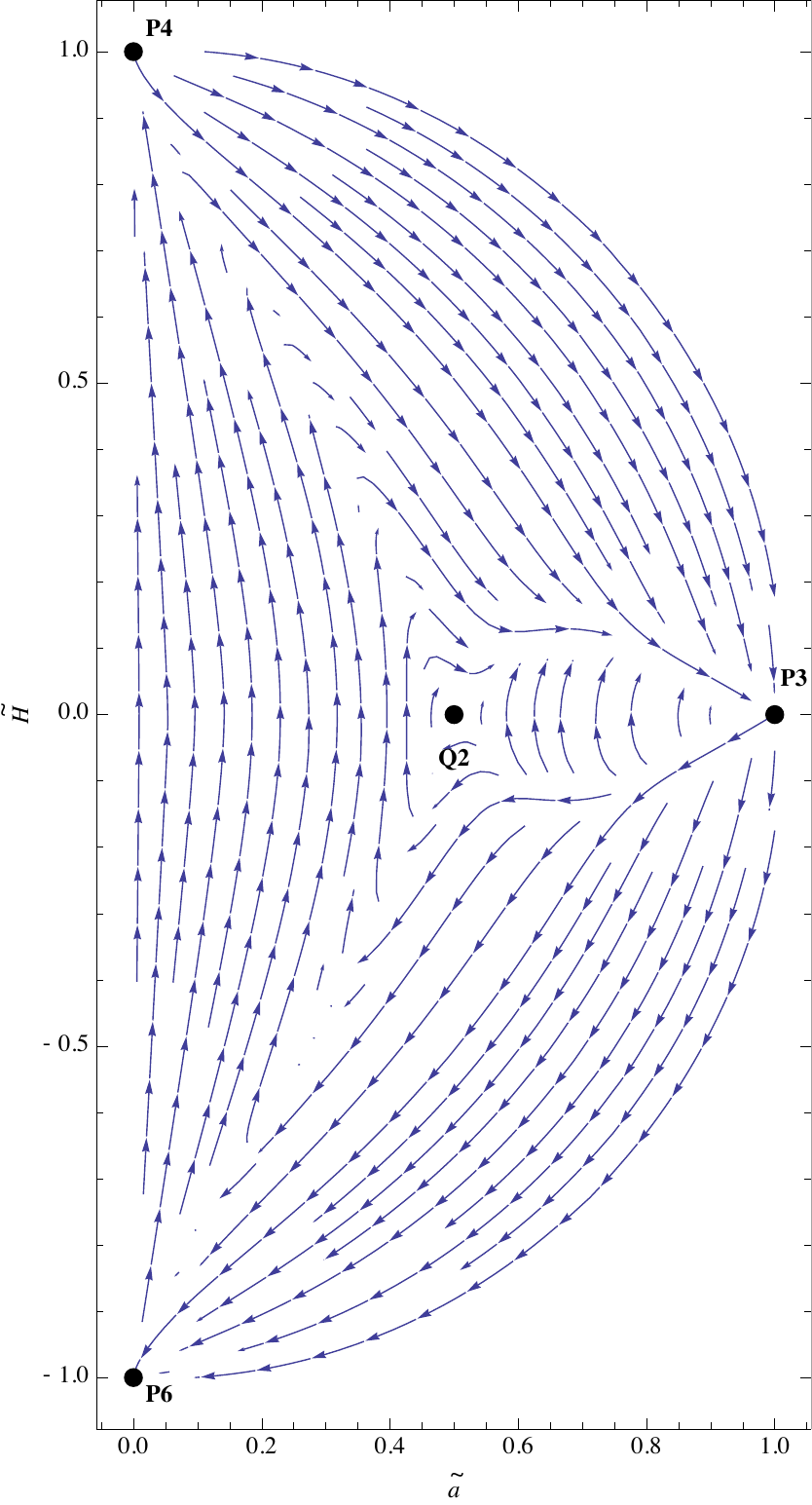}
\caption{Projected phase space of generalized HL cosmology with  $w=1/3$ and $\sigma_4=-k^2/(3\Lambda^2)$.}
\end{center}
\end{figure}
\begin{figure}\label{xx}
\begin{center}
\includegraphics[height=100mm]{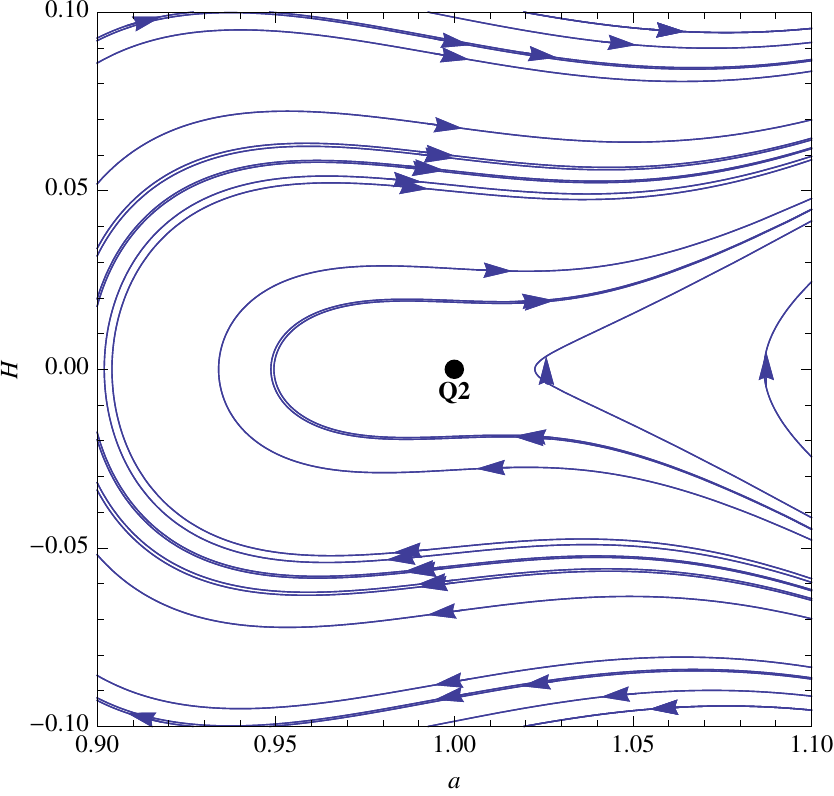}
\caption{Phase trajectories around the non-hyperbolic critical point $Q_2$.}
\end{center}
\end{figure}
Density $\rho$ is following:
\bea
\rho&=&-3 \Lambda (5+4 \Lambda\sigma_3)\ \ \textrm{at $Q_{1}$},\\
\rho&=&-3 \Lambda (-1 + \Lambda \sigma_3)\ \ \textrm{at $Q_{2}$}.
\eea
Density at $Q_{1}$ is positive if 
$$\{(\Lambda<0,\sigma_3<0); (\Lambda>0, 5/(4\Lambda)>\sigma_3)\}.$$
At $Q_{2}$, $\rho$ is positive for
$$\{(\Lambda<0, \Lambda\sigma_3>1); (\Lambda>0,\sigma_3<1/\Lambda),$$
plus conditions for existence: $Q_1$ has physical meaning if $k/\Lambda<0$ and $Q_2$ if $k/\Lambda>0$.

\subsection{General case}
In general critical points of the system  (\ref{pe2}) and (\ref{doth})  are of the following form:  $(a_x,0)$, where $a_x^2$ is a root of the equation:
 \be
 \Lambda(1 + w)x^3 -  { k(1+3w)}{ x^2} + {\sigma_3 (-1 + 3 w)k^2}{ x}
+{ 3\sigma_4 (-1 +  w)k}=0\label{bic1}.
\ee
Depending on the sign of the discriminant $\Delta=18a_0a_1a_2a_3-4a_2^3a_0+a_2^2a_1^2-4a_3a_1^3-27a_3^2a_0^3$, where $a_0=\Lambda(1 + w)$, $a_1= -  { k(1+3w)}$, $a_2={\sigma_3 (-1 + 3 w)k^2}$, $a_2={ 3\sigma_4 (-1 +  w)k}$, the above equation has one, two or three real solutions. Namely, for $\Delta>0$, there are three real roots, for $\Delta<0$ there is one real root and two complex conjugates, for $\Delta=0$  those conjugates become a real double root.

The eigenvalues of Jacobian matrix at those critical points read as:
\be
\left(- \sqrt{\frac{2D_1k}{a_x^6 (3 \lambda-1)}},\sqrt{\frac{2D_1k}{a_x^6 (3\lambda-1)}}\right),
\ee
where
\be
D_1= (1+3 w)a^4+2 k \sigma_3 (1-3 w) a^2-9 \sigma_4 (w-1).
\ee
For $w<-1/3$ and $-\sigma_4>\sigma_3^2 (1-3w)^2/((3w+1)(w-1)$ expression $D_1$ is always negative and it is always positive for $w>-1/3$ and $\sigma_4>\sigma_3^2 (1-3w)^2/((3w+1)(1-w)$. Thus depending on the sign of $k/(3\lambda-1)$ the critical points, if exist, are  either always stable or always unstable. Nature of their stability depends on the values of $a_x$, $\Lambda$, $\sigma_3$ and $\sigma_4$.

Stability properties of critical points at infinity is the same as for the detailed balance case. After the Poincar\`e transformation  (\ref{poin1}) and (\ref{poin2})
the whole phase space  is contained within a semi-circle ($a\ge0$) of radius one.

Points at $r=1$ and $\phi=\pi/2,\  3\pi/2$ are repelling  and attracting  node, respectively. Point $r=1$, $\phi=0$ is non-hyperbolic, and its stability properties can be obtained e.g. from numerical simulations.

Figure 6. shows the phase space of system with three  finite critical points. Points $S_{1}$ and $S_{3}$ are centers, point $S_{2}$ is a saddle.
\begin{figure}[!h]
\begin{center}
\includegraphics[height=100mm]{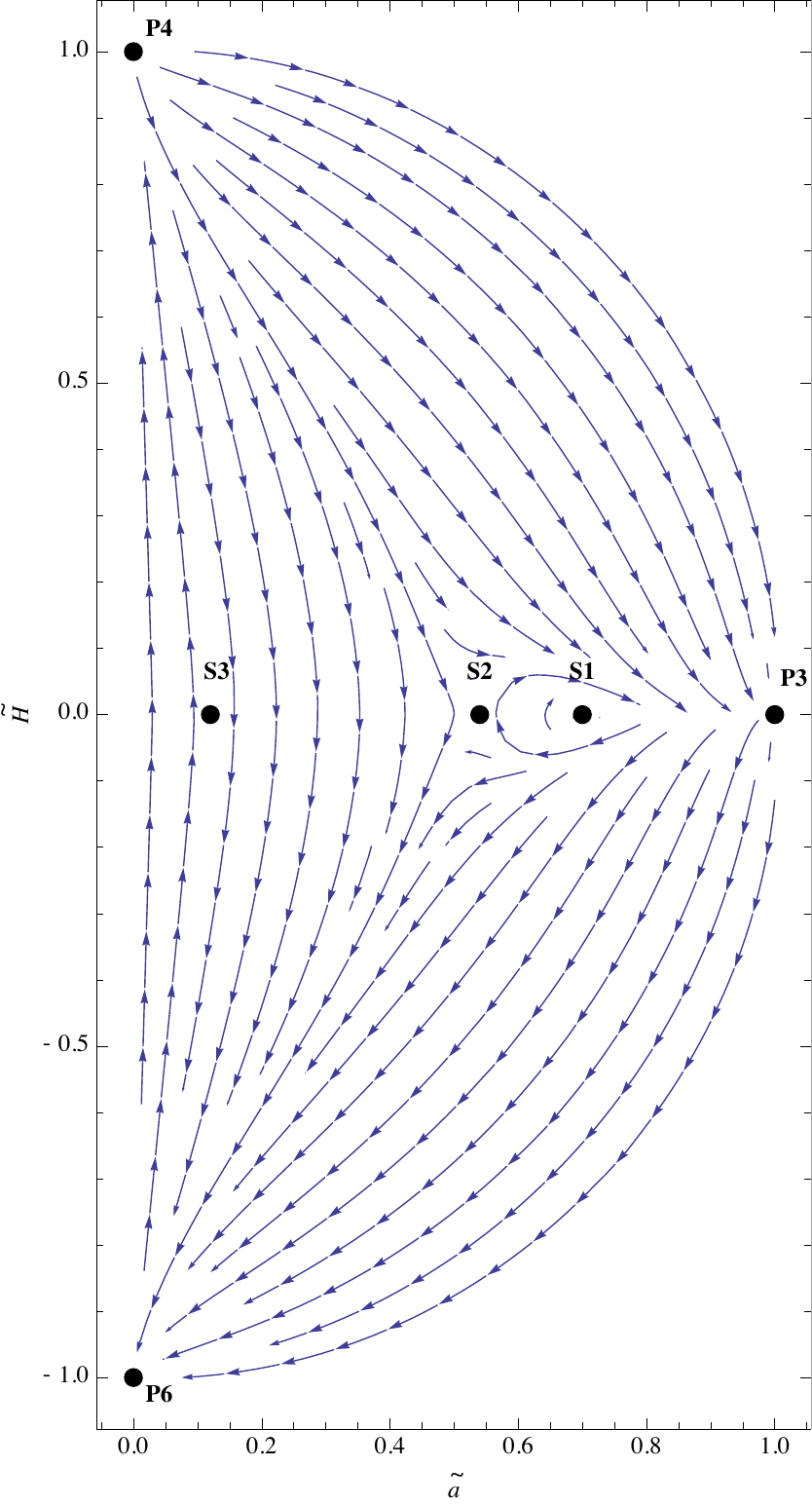}
\caption{Projected phase space of the generalized HL universe with 3 critical points existing}
\end{center}
\end{figure}

In Table 2. we have gathered properties of the finite critical points in the SVW generalization of the Ho\v{r}ava cosmology.\\

\begin{table}[!h]
\begin{center}
\begin{tabular}{|c|c|c|c|c|}
\hline
Point &    $w$ & Existence &Stability &$\rho$ positive   \\
\hline
\hline
 $P_{1}$ &-1&$(k\sigma_3>0,\sigma_4<0)$& center for&$k<0, \sigma_3<0$\\
&&& $k/(3\lambda-1)>0$&\\
 & & & saddle for &$0>\Lambda>1/\sigma_3$\\
  &&&$k/(3\lambda-1)<0$&\\
    \hline
 $P_{2}$&-1&$(k\sigma_3>0,$& center for &$k<0, \sigma_3<0$\\
 &&$0>\sigma_4>-\sigma_3^2/3)$& $k/(3\lambda-1)<0$&\\
 \cline{3-3}
 & & $(k\sigma_3<0,\sigma_4>0)$& saddle for  &$\Lambda<0$\\
 &&& $k/(3\lambda-1)>0$&\\
   \hline  
 $Q_{1}$&1/3&$(\sigma_4=-k^2/{3\Lambda^2},$& center for &$(\Lambda<0,\sigma_3<0)$\\
 && $k/\Lambda<0$)& $\Lambda/(3\lambda-1)<0$&\\
 & && saddle for &$ (\Lambda>0, \sigma_3<-5/(4\Lambda))$\\
 &&& $\Lambda/(3\lambda-1)>0$&\\
    \hline 
$Q_{2}$&1/3&$(\sigma_4=-k^2/{3\Lambda^2},$& cusp &$(\Lambda<0, \Lambda\sigma_3>1)$\\ 
 & & $k/\Lambda>0$)& &$( \Lambda>0,\sigma_3<1/\Lambda)$\\
    \hline 
$S_{1}$&$(-1,-1/3)$&depending on & $\sigma_4<-\frac{\sigma_3^2 (1-3w)^2}{(3w+1)(w-1)}$ &\\
$S_{2}$&& the sign of &  center for   &\\
$S_{3}$&&$\Delta$ and of the roots:& $k/(3\lambda-1)>0$&\\
&&0,1,2 or 3 points&saddle for&\\
&&&$k/(3\lambda-1)<0$&\\
\hline
$S_{1}$&$(-1/3,1/3)$&depending on &$\sigma_4>\frac{\sigma_3^2 (1-3w)^2}{(3w+1)(1-w)}$&\\
$S_{2}$&  $\cup (1/3,1])$& the sign of & center for   &\\
 $S_{3}$&&$\Delta$ and of the roots:& $k/(3\lambda-1)<0$&\\
&&0,1,2 or 3  points&saddle for&\\
&&&$k/(3\lambda-1)>0$&\\
\hline
   \end{tabular}\caption{Properties of the finite critical points in the SVW HL theory.} 
\end{center}   
\end{table}

\section{Conclusions}

In this work we have performed  a detailed analysis of a phase structure of the HL cosmology with and without detailed balance condition. 
Both this models contain a dark radiation term $1/a^4$ in the analogs of the Friedmann equations. Thus it is possible for a nonflat universe ($k\neq0$) that the Hubble parameter $H=0$ at some moment of time, which is a necessary  condition for the realization of the bounce. Comparing phase trajectories  obtained in those models we have attempted to answer the question how the generalization of Ho\v{r}ava gravity (breaking the detailed balance condition) impacts the occurrence and behavior of bouncing solutions. Additional term $1/a^6$ that appears in the Friedmann equations of SVW model, is of either sign, and thus it may possibly compensate the $1/a^4$ term (generic for HL gravity) leading to the singular solution.

Indeed, it occurred that the biggest difference between the Ho\v{r}ava theory and its generalization arrives for the small values of a scale parameter  $a$ and a Hubble parameter $H$. This is not surprising, as  the SVW gravity term $1/a^6$  plays role only  for the small values of $a$ and becomes insignificant for the bigger ones.

In the original Ho\v{r}ava formulation there may be two finite critical points, one of them a center and one a saddle. They are of the type $(a_x,0)$ in the $(a,H)$ space, thus strictly connected to  bounce solutions. These pairs of points exist both only for matter with $w>1/3$. Around a center there are closed orbits corresponding to the oscillating universe, i.e. going through eternal cycles of contraction, bounce and expansion. These orbits resemble bounce solution described by Brandenberger \cite{Brandenberger:2009yt} or quasi-stationary solutions presented in our previous work \cite{ewa} based on a field approach. Such solutions are physically interesting (density $\rho>0$) for $w<1/3$ and $k/\Lambda<0$ -- so either for a closed universe with a positive cosmological constant, or an open universe with $k=1$ and a negative cosmological constant $\Lambda$. The second class of oscillating solutions, with vanishing density $\rho=0$, appears when $k/\Lambda>0$. Additionally, there is a third bounce scenario around a linear center $P_{2}$ (moved to $\infty$ and coinciding with $P_{3}$). Here there are onefold closed trajectories: universe starts from a static one ($H=0$) and infinite ($a=\infty$), goes through a period of contraction to a finite size, then after a bounce starts expanding and again ends as a static infinite universe. Moreover, for some values of parameters, i.e. $k/\Lambda<0$ and $w>1/3$ there are no finite critical points, thus no bouncing solutions.

In the SVW HL cosmology,  with additional term appearing in the analogs of Friedmann equations,  there may exist 0,1,2 or 3 finite critical points. They are also of the type $(a_x,0)$ in $(a,H)$ space. Here $a_x$ is the solution of the bicubic equation. In general there exists at least one real solution of the cubic equation with real coefficients, but physical points correspond only to positive values of these roots.  Critical points might be stable centers -- surrounded by closed orbits, describing oscillating universes, or unstable saddles. There  also exist solutions with orbits around a  linear center  at $(\infty,0)$, where similarly as in the original HL theory, a universe starts from a static infinite the, collapses to a finite size, undergoes a bounce and then expands to a static infinite state. Thus there is one cycle only, without further oscillations. There are also sets of parameters, much wider than in the original HL theory, that do not allow the existence of finite critical points, leading  only to singular solutions.

The most significant feature of oscillating (and bouncing) solutions in the SVW formulation is the existence of two  centers, with a saddle between them (three finite critical points) for some values of parameters. We present such a solution at the Fig. 6. In a more realistic situation, that includes dynamical change of state parameter, it would be possible to go from one oscillating  bouncing solution to another. Present framework does not allow such evolution as it describes matter as  hydrodynamical fluid with a constant $w$. We expect that the field approach, with a more complete dynamics, may be suitable for further investigation of this interesting scenario.

The phase structure at infinity is the same at both formulations. Except bouncing solutions around finite critical points, there are also solutions leading to Big Bang, Big Crunch or eternal expansion. It is worth to stress that in both models, the original HL gravity and the SVW generalization, there are classes of parameters that do not allow a non singular evolution.
Physical interpretation of some of these parameters (coupling constants $\sigma_3$ and $\sigma_4$ in SVW model) still remains an open question.

\begin{center}
{\bf Acknowledgements}
\end{center}
This work has been
supported by the Polish Ministry of Science and Higher Education
grant   PBZ/MNiSW/07/2006/37.


\section*{References}

\begin{thebibliography}{00}

\bibitem{Ekp1}
Khoury J, Ovrut B A, Steinhardt  P J and Turok N 2001
%``The ekpyrotic universe: Colliding branes and the origin of the hot big  bang,''
{\it Phys. Rev.}  D {\bf 64}  123522 
[arXiv:hep-th/0103239].
%%CITATION = HEP-TH 0103239;%%,
\bibitem{Ekp2} Buchbinder E  I, Khoury  J  and  Ovrut B  A 2007 {\it Phys. Rev.} D {\bf 76} 123503     [arXiv:hep-th/0702154].

\bibitem{Cyclic1}
Steinhardt P J and Turok  N 2002
%``A cyclic model of the universe,''
{\it Science}  {\bf 296.} no. 5572  1436  [arXiv:hep-th/0111030].
\bibitem{Cyclic2}
Steinhardt P J and Turok  N 2002
% ``Cosmic evolution in a cyclic universe,''
{\it Phys. Rev.} D  {\bf 65} 126003  [arXiv:hep-th/0111098].
\bibitem{Cyclic3}
Steinhardt PJ and Turok  N 2002
%``Is Vacuum Decay Significant in Ekpyrotic and Cyclic Models?''
{\it Phys. Rev.} D  {\bf 66} 101302  [astro-ph/0112537].



\bibitem{pyrotech}
Kallosh R, Kofman  L and Linde  A D 2001
{\it Phys. Rev.} D  {\bf 64} 123523 
% ``Pyrotechnic universe,''
[arXiv:hep-th/0104073].

\bibitem{ap1}
Ashtekar A, Pawlowski T and Singh P   2006
%Quantum Nature of the Big Bang
{\it Phys.~Rev.~Lett.} {\bf 96} 14130
[arXiv:gr-qc/0602086].
\bibitem{ap2}
Ashtekar A, Pawlowski T and Singh P 2006
%Quantum Nature of the Big Bang: An Analytical and Numerical Investigation
{\it Phys.~Rev.} D~ {\bf 73}~124038 
[arXiv:gr-qc/0604013].

\bibitem{ap3}
Ashtekar A, Pawlowski T and Singh P 2006
%Quantum Nature of the Big Bang: Improved dynamics
{\it Phys.~Rev.} ~D~ {\bf 74} 084003 
[arXiv:gr-qc/0607039].

\bibitem{bw}
Maartens R 2004 {\it Living Rev. Rel.}  {\bf 7} 7 [arXiv:gr-qc/0312059].

\bibitem{rs}
 Randall L and  Sundrum  R 1999 {\it Phys. Rev. Lett. } {\bf 83} 4690 [arXiv:hep-th/9906064].


\bibitem{bc}
Novello M and Perez S E 2008
%Bouncing Cosmologies,
{\it Phys. Rept.}  {\bf 463} 127-213 
    [arXiv:0802.1634 [astro-ph]].



\bibitem{hor1}
  Horava  P  2009
  %{\em ``Membranes at Quantum Criticality,''}
  {\it JHEP} {\bf 0903} 020
  [arXiv:0812.4287 [hep-th]].
  %%CITATION = JHEPA,0903,020;%%

\bibitem{horava}
   Horava  P  2009
  %{\em ``Quantum Gravity at a Lifshitz Point,''}
  {\it Phys. Rev.}  D {\bf 79}  084008
  [arXiv:0901.3775 [hep-th]].
  %%CITATION = PHRVA,D79,084008;%%

%\cite{Horava:2009if}
\bibitem{Horava:2009if}
  Horava  P 2009 {\it Phys. Rev. Lett.} {\bf102} 161301
 % {\em ``Spectral Dimension of the Universe in Quantum Gravity at a Lifshitz
 % Point,''}
  [arXiv:0902.3657 [hep-th]].
  %%CITATION = ARXIV:0902.3657;%%

%\cite{Nastase:2009nk}
\bibitem{Nastase:2009nk}
  Nastase H
  % {\em ``On IR solutions in Horava gravity theories,''}
 [arXiv:0904.3604 [hep-th]].
  %%CITATION = ARXIV:0904.3604;%%
  
\bibitem{sfetsos}
Kehagias A and  Sfetsos K 2009 {\it Phys. Lett.} B {\bf 678} 123 	[arXiv:0905.0477 [hep-th]].

%\cite{Sotiriou:2009}
\bibitem{Sotiriou:2009}  
  Sotiriou T P, Visser M and Weinfurtner S 2009 {\it Phys. Rev. Lett.} {\bf102} 251601 [arXiv:0904.4464 [hep-th]].


\bibitem{SVW}
 Sotiriou T P, Visser M and Weinfurtner S 2009
{\it JHEP} {\bf 0910} 033, [arXiv:0905.2798 [hep-th]].
 
\bibitem{mk}
Mukohyama S 2010 {\it invited review for Class. Quantum Grav.} [arXiv:1007.5199[hep-th]].




%\cite{Charmousis:2009tc}
\bibitem{Charmousis:2009tc}
  Charmousis C, Niz G, Padilla  A and Saffin P M  2009
 %{\em ``Strong coupling in Horava gravity,''}
{\it JHEP}  {\bf 0908} 070  [arXiv:0905.2579 [hep-th]].
  %%CITATION = ARXIV:0905.2579;%%


\bibitem{LiPang}
 Li M and  Pang Y 2009
%" A Trouble with Horava-Lifshitz Gravity'',
{\it JHEP} {\bf 0908} 015  [arxiv: 0905.2751[hep-th]].





%\cite{Bogdanos:2009uj}
\bibitem{Bogdanos:2009uj}
Bogdanos C and Saridakis  E N 2010
%Perturbative instabilities in Horava gravity
{\it Class. Quantum Grav.}  {\bf 27} 075005  
[arXiv:0907.1636].



\bibitem{ak}
Kobakhidze A 2010 {\it Phys. Rev.} D {\bf 82} 064011 	[arXiv:0906.5401 [hep-th]].



\bibitem{sc1}
Iengo R, Russo J G and Serone M 2009 {\it JHEP} {\bf 0911} 020 [arXiv:0906.3477 [hep-th]].

\bibitem{sc2}
 Cai R G, Hu B and  Zhang H B 2009 {\it Phys. Rev.} D {\bf 80}
041501  [arXiv:0905.0255 [hep-th]].


\bibitem{blas}
Blas D, Pujolas O and  Sibiryakov S 2009 {\it JHEP} {\bf 091} 029 [arXiv:0906.3046 [hep-th]].

\bibitem{blas3}
Blas D, Pujolas O and  Sibiryakov S 2010 {\it Phys. Lett. B} {\bf 688} 350 [arXiv:0912.0550 [hep-th]] 



\bibitem{hc1}
  G.~Calcagni 2009
% {\em ``Cosmology of the Lifshitz universe,''}
{\it JHEP}  {\bf 0909} 112   [arXiv:0904.0829 [hep-th]].
  %%CITATION = ARXIV:0904.0829;%%


\bibitem{hc2}
  Kiritsis E and Kofinas G  2009
 %{\em ``Horava-Lifshitz Cosmology,''}
{\it Nucl. Phys.} B  {\bf 821} 467   [arXiv:0904.1334 [hep-th]].
  %%CITATION = ARXIV:0904.1334;%%

% \cite{Saridakis:2009bv}
\bibitem{Saridakis:2009bv}
 Saridakis  E N 2010
 {\it Eur. Phys. J.} C {\bf 67} 
  %``Horava-Lifshitz Dark Energy,''
  [arXiv:0905.3532 [hep-th]].
  %%CITATION = ARXIV:0905.3532;%%
  


%\cite{Brandenberger:2009yt}
\bibitem{Brandenberger:2009yt}
  Brandenberger R 2009 {\it Phys. Rev.} D {\bf 80} 043516  % {\em ``Matter Bounce in Horava-Lifshitz Cosmology,''}
 % {\em ``Matter Bounce in Horava-Lifshitz Cosmology,''}
  [arXiv:0904.2835 [hep-th]].
  %%CITATION = ARXIV:0904.2835;%%




%\cite{Mukohyama:2009zs}
\bibitem{Mukohyama:2009zs}
  Mukohyama S, Nakayama K, Takahashi  F and Yokoyama S 2009
 {\it Phys. Lett.} B {\bf 679} 6
 % {\em ``Phenomenological Aspects of Horava-Lifshitz Cosmology,''}
  [arXiv:0905.0055 [hep-th]].
  %%CITATION = ARXIV:0905.0055;%%


%\cite{Minamitsuji:2009ii}
\bibitem{Minamitsuji:2009ii}
 Minamitsuji  M 2010
 {\it	Phys. Lett.} B {\bf 684} 194
 %{\em``Classification of cosmology with arbitrary matter in the
 % Ho\v{r}ava-Lifshitz theory,''}
  [arXiv:0905.3892 [astro-ph]].
  %%CITATION = ARXIV:0905.3892;%%

%\cite{Wang:2009rw}
\bibitem{Wang:2009rw}
 Wang  A and Wu Y  2009
% {\em ``Thermodynamics and classification of cosmological models in the
 % Horava-Lifshitz theory of gravity,''}
{\it JCAP} {\bf 0907} 012  [arXiv:0905.4117 [hep-th]].
  %%CITATION = ARXIV:0905.4117;%%

%\cite{Takahashi:2009wc}
\bibitem{Takahashi:2009wc}
 Takahashi T and Soda J 2009
 %``Chiral Primordial Gravitational Waves from a Lifshitz Point,''
{\it  Phys.\ Rev.\ Lett.}   {\bf 102} 231301
 [arXiv:0904.0554 [hep-th]].

\bibitem{lmp}
Lu H,   Mei H  and Pope C N 2009
{\it Phys. Rev. Lett.} {\bf 103} 091301
%?Solutions to Horava Gravity,? 
[arXiv:0904.1595 [hep-th]].

\bibitem{gh}
 Ghodsi A and  Hatefi E 2010 {\bf Phys. Rev.} D {\bf 81} 044016 [arXiv:0906.1237 [hep-th]].

\bibitem{SonKim}
 Son E J and  Kim W
[arXiv:1007.5371 [gr-qc]].


\bibitem{Saridakis:2009}
 Cai Y-F and  Saridakis E N 2009 {\it JCAP} {\bf 0910} 020 
[arXiv:0906.1789 [hep-th]].

%\cite{Gao:2009wn}
\bibitem{Gao:2009wn}
 Gao X, Wang Y, Xue W and Brandenberger R 2010
 %``Fluctuations in a Ho\v{r}ava-Lifshitz Bouncing Cosmology,''
{\it JCAP} {\bf 1002}  020 
 [arXiv:0911.3196 [hep-th]].
 %%CITATION = JCAPA,1002,020;%%










\bibitem{ewa}
Czuchry E 2011 {\it Class. Quantum Grav.}  {\bf 8}  085011 [arXiv:0911.3891 [hep-th]].

\bibitem{ndb1}
Wu P and  Yu H 2010 {\it Phys. Rev.} D {\bf 81}   103522 
	[arXiv:0909.2821 [gr-qc]].

\bibitem{ndb2}
 Maeda K,  Misonoh Y and  Kobayashi T 2010 {\it Phys. Rev.} D {\bf 81} 103522
 	[arXiv:1006.2739 [hep-th]].




\bibitem{ps1}
Carloni S, Elizalde E and Silva P J {\it  Class. Quantum Grav.} {\bf 27} 045004 
%An analysis of the phase space of Horava-Lifshitz cosmologies
[arXiv:0909.2219 [hep-th]].


\bibitem{ps2}
Leon G and Saridakis E N 2009  {\it JCAP} {\bf 0911} 006
%Phase-space analysis of Horava-Lifshitz cosmology
{\it JCAP} {\bf 0911} 006 [arXiv:0909.3571 [hep-th]].


\bibitem{BDEL}
 Bin\'{e}truy  P, Deffayet  C,  Ellwanger U and  Langlois D 2000
%\textit{Brane cosmological evolution in a bulk with cosmological constant},
{\it  Phys.\ Lett.}  B  {\bf 477} 285  [hep-th/9910219].


\bibitem{xx}
Arrowsmith D K  and Place C M 1990 {\it An introduc-
tion to Dynamical Systems} Cambridge University Press.



%\cite{frolov}
\bibitem{frolov}
 Felder G N,  Frolov A and Kofman L   2002
%Warped Geometry of Brane Worlds
{\it Class. Quantum Grav.}  {\bf 19} 2983
 [arXiv:hep-th/0112165].

\bibitem{frolov2}
 Felder G N,  Frolov A, Kofman L and  Linde A 2002
{\it Phys.~Rev. }D  {\bf 66} 023507     [arXiv:hep-th/0202017].








\end{thebibliography}
\end{document}